# Strong plasmon coupling between two gold nanospheres on a gold slab

## H. Liu[1, *], J. Ng[2], S. B. Wang[2], Z. H. Hang[2], C. T. Chan[2] and S. N. Zhu[1]

[1]National Laboratory of Solid State Microstructures and Department of Physics, Nanjing University, Nanjing, 210093, China

[2]Department of Physics and Nano Science and Technology Program, Hong Kong University of Science and Technology, Clearwater Bay, Hong Kong, China

E-mail: *liuhui@nju.edu.cn; URL：http://dsl.nju.edu.cn/mpp

**Abstract**: In this work, the plasmon coupling effect between two gold nanospheres on a gold slab is investigated. At plasmon resonance frequencies, electrons on the surface of the slab are absorbed into spheres and contribute to plasmon oscillation. This effect can help enhance the local electric field and optical coupling force between the two spheres.

Our recent papers about **"force induced by plasmonic coupling"**:

[1] Strong Light-Induced Negative Optical Pressure Arising from Kinetic Energy of Conduction Electrons in Plasmon-Type Cavities

**Phys. Rev. Lett.** 106, 087401 (2011)

[2] Sizable microwave force in parallel-plate metallic cavity

**Phys. Rev. B** (2011) accepted

## 1. Introduction

In recent years, surface plasmon resonances (SPR) in various metallic nanostructures have been widely reported. SPRs of nanostructures are applied in surface enhanced Raman scattering (SERS) and other nonlinear optical processes because of their strong local electric field enhancement effect. For complex nanosystems composed of many nanostructures, various plasmonic coupling effects between these structures are very interesting and useful [1-6]. The near field coupling between these nanostructures could produce "hot spots" and further enhance the local field. Many coupled plasmonic structures have been investigated thus far, including nanospheres [7, 8], nanoshells [9], nanorods [10-13], split-ring resonators [14-16], multilayer of fish-nets [17], and slit-hole resonators [18]. Plasmonic coupling is also able to produce a strong optical force between the nanostructures. Such forces between nanostructures could affect the signal of the surface-enhanced Raman spectroscopy, or result in aggregation of nanosystems [19-23].

In this paper, the plasmon coupling between two gold nanospheres on a gold slab is investigated. Two different resonance modes could be established in this system because of the induced current on the surface of the slab. The exchange current between the two spheres further enhance the local electric field in the structure, resulting in a strong plasmon coupling force between the two spheres.

## 2. Structure and numerical model

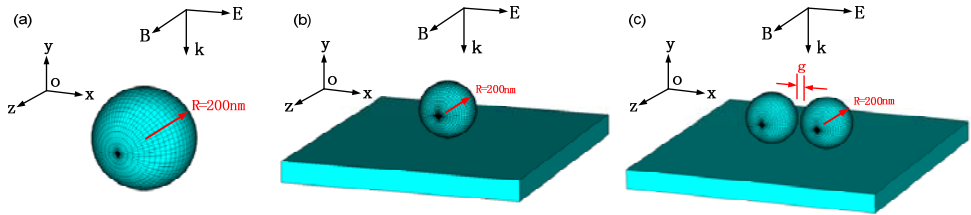

**Figure 1**. (a) single sphere without slab; (b) single sphere on slab; (c) two spheres on slab. The incident plane wave propagates along (-y) direction with its polarization in x direction. Open boundaries are used in all directions.

Figure 1 presents the geometry of several structures investigated in this paper, together with the design parameters. Figure 1 (a) is one gold sphere in the air. In Figure 1 (b), the gold sphere is placed on a gold slab. Figure 1 (c) shows two identical gold spheres placed on a gold slab. The radius of each of the three structures is R=200 nm. The size of the slab is 2000×2000 nm with a thickness of 150 nm. The gap between the two spheres in Figure 1 (c) is defined as g, which is tuned from 10 to 75 nm in this work. In this work, the sphere of this size are chosen because it is very simple nanostructures and widely found in experiments. Besides, its resonance is not too strong to overtax the simulation software. In all three configurations, the incident EM wave propagates in the (-y) direction with its polarization in the x direction. Open boundaries are used in all the directions.

A full-wave simulation using time-domain simulation software (CST Microwave Studio) is used to study the EM response of the structures. In the simulation, the metal permittivity is given by $\varepsilon(\omega) = 1 - \omega_p^2 / (\omega^2 + i\omega_\tau \omega)$, where $\omega_p$ is the bulk plasma frequency and $\omega_p$ is the relaxation rate. For the gold, the characteristic frequencies fitted to experimental data are $\omega_p = 2.175 \times 10^3 \text{THz}$ and $\omega_\tau = 6.5 \text{THz}$.

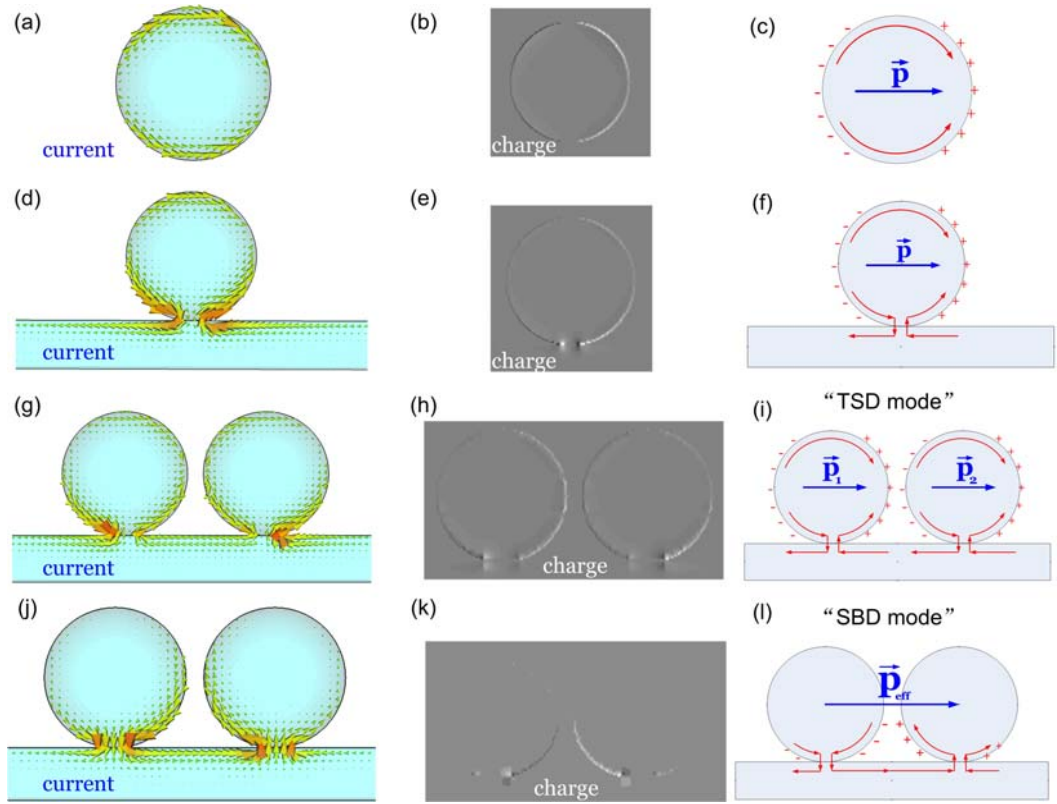

**Figure 2.** Distribution of induced current density and charges of four resonance modes:

(a-b) one sphere without slab (denoted as single electric dipole in (c)); (d-e) one sphere on slab (denoted as single electric dipole in (f)); (g-h) higher frequency mode of two spheres on slab (denoted as two small dipoles mode (TSD) in (i)); (j-k) lower frequency mode of two spheres on slab(denoted as single big dipole mode (SBD) in (l)). Here, the charge distributions (in (b), (e), (h), (k)) are given as grayscale maps with white dots being positive charge and dark dots being negative charge.

**3. Two plasmonic resonance modes**

First, one gold sphere (R=200 nm) without a slab is investigated. In the simulations, an EM wave incident to a plane is directed onto the sphere with its polarization along the x direction [Fig. 1 (a)]. In the program, the local electric field $E_x$ near the surface of the sphere is recorded. In principle, if we could determine the local e field and conductivity of material, we could obtain the current density based on equation $\vec{j} = \sigma \cdot \vec{E}$. Here, the metal is defined as drude model and its σ could be calculated. E could be obtained directly in the simulations, and then current density could be calculated. Excited by the incident wave, electrons oscillate in the sphere. This produces plasmon resonance. In the simulation, when the incident frequency is swept from 0 to 400 THz, the recorded electric field shows one resonance peak at 182 THz. At this frequency, the distribution of the induced current density inside the sphere $\vec{j}$ is obtained from the simulation, as given in Fig. 2 (a). Based on continuity equation of charges $\nabla \cdot \vec{j} + \frac{\partial \rho}{\partial t} = 0$, the distribution of charge density is able to obtained as $\rho = \frac{1}{i\omega} \nabla \cdot \vec{j}$. For this resonance mode, positive and negative charges accumulate at the two sides of sphere [Fig.2 (b)]. This makes the sphere behave similarly to an electric dipole [Fig. 2 (c)]. For a quite small nanosphere, all the electrons inside the structure could be excited and attribute to the plasmon oscillation. Such a mode is volume plasmon resonance mode. In our structure, the diameter of sphere is about 400nm. The skin depth of induced current is only about 20nm. Then the plasmon mode in nanospheres of this size is a kind of surface modes. For such a surface mode, its eigen frequencies depend on its geometry size.

Next, one sphere on gold slab is investigated [Fig. 1 (b)]. The size of the slab is much larger than the sphere; thus the influence of the boundary of the slab can be ignored. For the local field near the surface of sphere, only one resonance peak is found at 227 THz. The current density at this resonance mode is given in Fig. 2 (d). The results show that when the sphere is placed on the surface of the metal slab, the electrons on the surface of slab are absorbed into the sphere, contributing to the plasmon resonance. As more electrons accumulate on the two sides of sphere [Fig.2 (e)], the plasmon resonance becomes stronger, and better local field enhancement is produced. The sphere on the slab can still be considered an electric dipole [Fig. 2 (f)].

The properties of the two spheres on the slab [Fig. 1 (c)] are investigated. In the program, the x-component local electric field $E_x$ in the gap between the two spheres is recorded. In our simulations, the gap between the two spheres is tuned from 10 to 75 nm. When the incident frequency is swept from 0 to 400 THz, the recorded electric field, $E_x$, is obtained in the simulations. This is shown in Fig. 3 (a). For the structure with a different gap, g, the recorded E fields are given as the different curves in the figure. For all the curves, two resonance peaks are found in the range of 0-400 THz. The lower resonance frequency is denoted as $\omega_1$, and the high resonance frequency is denoted as $\omega_2$. The current densities at $\omega_1$ and $\omega_2$ are given in Figs. 2 (j) and (g) to investigate the differences of these two resonance modes. For both modes, electrons on the slab surface join the plasmon resonance. At high frequency $\omega_2$ [Fig. 2 (e)], most of electrons in each sphere continue to oscillate within each sphere and contribute to the plasmon resonance mode of two separated spheres. Positive and negative charges accumulate on the two sides of each sphere [Fig.2 (h)]. The exchange of current between the two spheres remains quite small. This makes the two spheres behave as two separated electric dipoles [Fig. 2 (i)]. This high-frequency mode could be called the two small dipoles mode (TSD). For the low-frequency mode at $\omega_1$ [Fig. 2 (j)], the resonance process is quite different. Most of the oscillating electrons can no longer be confined in one sphere because of the help of the slab. They are able to flow from one sphere to the other. Most of electrons oscillate between two spheres because of this strong exchange in current effect. When positive charges accumulate in one of the spheres, negative charges accumulate in the other one [Fig. 2 (k)]. Thus, the two spheres cannot be seen as two separate dipoles, they must be seen as one big effective electric dipole [Fig. 2 (l)]. This low-frequency mode is called the single big dipole mode (SBD).

In the simulations, the local E field enhancement between the two spheres with a different gap is calculated [Fig. 3 (a)]. The gap size g is tuned from 10 to 75 nm. For the lower resonance mode, the oscillation of electrons is not confined inside one sphere and could flow from one sphere to the other. The whole system could be seen as U-shaped LC circuit. The low eigenfrequency is actually the LC resonance frequency $\omega = 1/\sqrt{LC}$. The gap between two spheres could be seen as the capacitance of the LC circuit. When gap g is increased, the capacitance $C \sim 1/g$ will decrease. As a result, the resonance frequency will increase. For comparison, the frequency dependence of the E field between the two spheres without a slab is also calculated. Gap values of 10, 20, and 30 nm are chosen. The recorded electric fields are given in Fig. 3 (b). For all the three curves, only one resonance peak was found (denoted as $\omega_0$). The E field between the two spheres is also increased when g is reduced. Comparing Figs. 3 (a) and 3 (b), for the same gap value, both resonance modes of the structure with the slab possess stronger E field enhancement than the structure without the slab. The induced current in the slab surface introduces more electrons into the plasmon resonance. This could help enhance local field of the sphere. A comparison of Figs. 3 (a) and 3 (b) verifies this analysis. The slab is helpful in increasing the local E field between the two spheres.

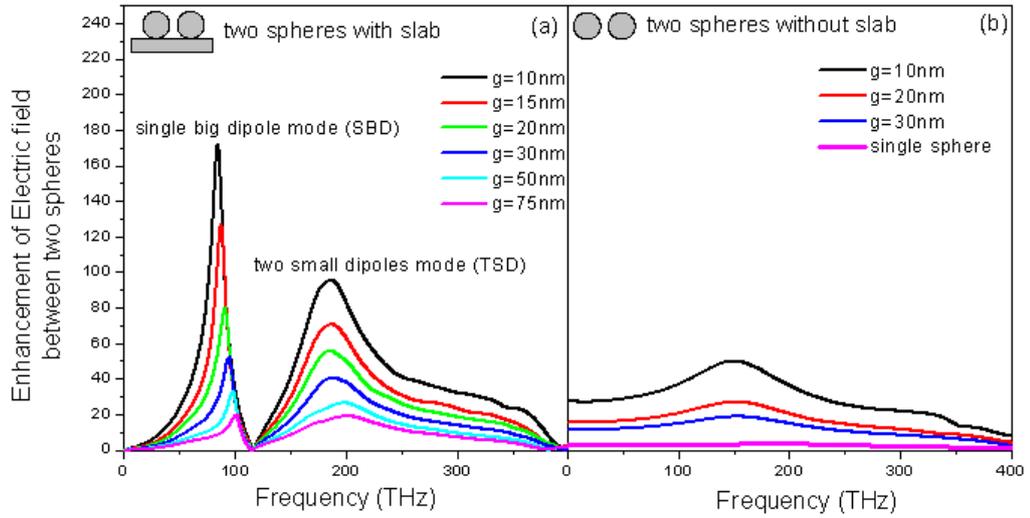

**Figure 3.** Enhancement of electric field between two spheres with slab (a) and without slab (b).

Aside from the comparison between the structure with slab and the structure without slab, the difference between the TSD mode and the SBD mode of the structure with slab [Fig. 3 (a)] is also compared. When the gap is reduced from 75 to 10 nm, both E fields of the two modes increase. However, the E field of the SBD mode $E(\omega_1)$ increases more rapidly than that of the TSD mode $E(\omega_2)$. When g=75 nm, the E field of the SBD mode is nearly equal to that of the TSD mode $E(\omega_1) \simeq E(\omega_2)$. When g is reduced, SBD becomes stronger than TSD. At g=10 nm, the E field of the SBD is approximately twice that of the E field of TSD $E(\omega_1) \simeq 2 \cdot E(\omega_2)$. For the TSD mode, most of the electrons remain confined in each sphere, and the field enhancement mainly comes from the addition of two separated small electric dipoles. For the SBD mode, a strong exchange current between the two spheres allows more electrons join the oscillation. More incident EM wave energy is absorbed into the energy of electrons. This makes the E field enhancement of the SBD mode bigger than that of the TSD mode. Actually, we have already calculated the structure in which spheres do not touch the slab. In the results, the two spheres cannot exchange electrons with each other. The strong local field enhancement disappears and resonance coupling is much weaker. These improves that the touching is very important.

**4. Optical force between two spheres on the slab**

Previous works [19-22] have shown that the plasmon coupling between two metal spheres could result in a strong coupling force between the two spheres. Our results show that the two gold spheres on a gold slab present a much stronger plasmon resonance and local field enhancement than the two spheres without the slab. The gold slab is then expected to help increase the optical coupling force between the two spheres.

The optical force between the two spheres on a slab and the optical force between the two spheres without a slab are compared. The calculation of the optical force is based on the Maxwell tensor method. In the simulations, both the electric field $E(r)$ and magnetic field $H(r)$ are obtained in the space around the structures. After obtaining the EM fields, the time-averaged optical force $\langle \mathcal{F} \rangle_t$ between the two spheres can then be calculated using the Maxwell's stress tensor via a surface integral $\langle \mathcal{F} \rangle_t = \oiint_S \langle T \rangle_t \cdot ds$, where $T_{\alpha\beta} = \varepsilon_0 \left( E_\alpha E_\beta - E \cdot E \delta_{\alpha\beta} / 2 \right) + \mu_0 \left( H_\alpha H_\beta - H \cdot H \delta_{\alpha\beta} / 2 \right)$, and S is the integration surface enclosing one sphere. In the process, only the optical force between the two spheres is calculated. As the two spheres are placed on the slab along the x direction and the incident E field is polarized along the x direction, the interacting optical force occurring in the middle should be along the x direction. Then, only the x component of the optical force $F_x$ is considered.

Firstly, we calculate the frequency dependence of the optical force between the two spheres on the slab with the gap g=50 nm and the incident intensity being $1.0 \, \text{mW} / \mu\text{m}^2$. The result is given in Fig. 4 (a). The simulation shows that the optical force between the two spheres is an attractive force. At these two plasmon resonance frequencies, the optical force is enhanced as anticipated. For the SBD mode at a low resonance frequency, the optical force is 67 pN, whereas for the TSD mode at a high resonance frequency, the optical force is merely 21 pN. The former is three times that of the latter. The exchange current is formed between the two spheres in the SBD mode. This exchange current causes a much stronger optical force in the SBD mode than that in the TSD mode. For a single sphere in air, the scattering force by a plane wave is about 0.8 pN at resonance frequency, which is smaller than SBD mode by two orders of magnitude.

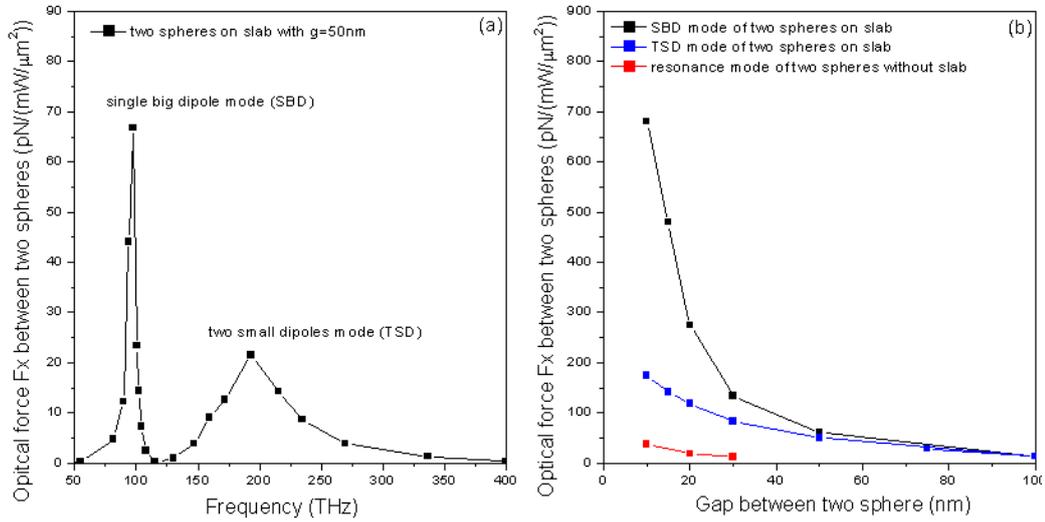

**Figure 4**. (a) The frequency dependence of optical force of two spheres on slab (with g=50nm); (b) The dependence of optical force on gap between two spheres with or without slab.

Second, the dependence of the attractive optical force on the gap between the two spheres is calculated. The results are given in Fig. 4 (b). Only optical forces at the two resonance frequencies are provided. For comparison, the results for the two spheres without the slab are given. The changes of optical force for all the three cases are stronger for a smaller gap. However, both modes of the two spheres on a slab possess stronger forces than the resonance mode of the two spheres without the slab (denoted as the red dotted curve). For a large gap value, the optical force of the SBD mode (denoted as the black dotted curve) is nearly equal to the TSD mode (denoted as the blue dotted curve). However, when the gap decreases, the optical force of the SBD mode increases much faster than the TSD mode. At g=10 nm, the optical force of the SBD mode is four times that of the TSD mode and eighteen times that of the case without the slab. These results prove that the exchange current induced in the slab helps increase the optical force between the two spheres greatly.

**5. Conclusions**

In summary, this work shows that, given nanospheres placed on a slab, electrons on the surface of slab are able to join in the plasmon oscillation and contribute to the resonance mode. This effect makes the plasmon resonance much stronger. Moreover, it produces much better local field enhancement. Two resonance modes are introduced in this system: the SBD mode at low frequency and the TSD mode at high frequency. Much stronger optical coupling force is produced by the SBD mode than the TSD mode because of the exchange current between the two spheres. Aside from the optical force, the slab could also be used to enhance the Raman scattering and other nonlinear optical effects in nanospheres in the future.

**Acknowledgments**

The research of H.L. was financially supported by the National Natural Science Foundation of China (No. 11074119, No.10874081, No.60907009, No.10904012), and by the National Key Projects for Basic Researches of China (No. 2010CB630703). The work in Hong Kong is supported by RGC Hong Kong through grant 600308 and by Nano Science and Technology Program of HKUST. Computation resources are supported by Hong Kong Shun Hing Education and Charity Fund.